\documentclass[12pt]{iopart}
\usepackage{setstack,iopams}
\usepackage{hyperref}
\usepackage{graphicx}
\usepackage{cite}
\hypersetup{colorlinks,linkcolor={blue},citecolor={blue},urlcolor={red}}
\begin{document}
\title[On dynamic aspects of the Proca field screening by a black hole]{On dynamic aspects of the Proca field screening by a black hole}

\author{Andrei L. Erofeev$^{1,2}$}

\address{$^1$ Budker Institute of Nuclear Physics, SB RAS, Novosibirsk, 630090, Russia}
\address{$^2$ Novosibirsk State University, Novosibirsk, 630090, Russia}
\ead{andrikola@yandex.ru}

\begin{abstract}
Classical black holes are known for almost half a century to nullify exterior classical massive vector field associated with a charge crossing the event horizon. This implies that, from the point of view of an external observer, the Proca field of the charge is screened with the strength gradually increasing as the charge adiabatically approaches the event horizon. In this paper we reject the adiabaticity constraint and calculate analytically the field evolution with respect to a distant observer in the frame of the simplest model of a contracting charged spherical shell concentrically surrounding a Schwarzschild black hole. 
We show that a time scale of the screening is determined by a mass of the black hole and, moreover, loss of Proca hair during the collapse of charged matter has the same temporal character. The latter conclusion disproves results of a number of papers.
Due to existence of the event horizon, there is discontinuous jump between massive and massless electrodynamics. This means that presence of an arbitrarily small mass of the photon gives rise to experimentally observed effects including generation of electric asymmetry of the Universe and galactic magnetic fields.
\end{abstract}
\section{Introduction}
The lack of massive vector fields associated with black holes (BH) is a particular conclusion of no-hair theorems proved by Bekenstein and Teitelboim. This statement was proved for both the spherically symmetric \cite{Bekenstein1,Teitelboim} and the stationary \cite{Bekenstein2} BHs. Moreover, an asymptotically flat spherically symmetric solution of the Einstein-Proca equations was shown by numerical calculations of Obukhov and Vlachynsky \cite{Obukhov} to be inevitably horizonless.

An analysis \cite{Leaute,Dolgov} demonstrated a non-trivial dependence of the Proca static field created by a uniformly charged spherical shell surrounding a Schwarzschild BH on the distance between the shell surface and the event horizon. The measured field from the point of view of an observer fixed outside the charged sphere gradually disappears as the surface of the sphere adiabatically approaches the horizon. This suppression of the field can be considered as effective screening of the charge.

A similar smooth behavior of the static field occurs for an arbitrarily small mass value of the gauge boson. In other words, the transition to Maxwell electrodynamics is discontinuous \emph{in the static case}. As shown in \cite{Teitelboim,Leaute,Dolgov,Vilenkin} this discontinuity is explained by the difference in the boundary conditions for the field at the horizon.

However, it is argued sometimes that the solution of a fully dynamical problem will manifest the continuous connection between the behavior of massive and massless fields in the vicinity of the BH. As stated without proof in the paper of Coleman, Preskill and Wilczek \cite{Coleman}, when the charge crosses the horizon, the Schwarzschild time scale for the field of the charge to vanish is consistent with the Compton wavelength of the gauge boson. 
Thus, as the mass of the boson gets smaller, the time scale for the field to decay gets longer and longer with Maxwell case realized in the limit of the mass going to zero.

An attempt to prove the statement \cite{Coleman} was made in \cite{Pawl}. The author of this paper concluded that the monopole massive vector field created by spherically symmetric distribution of matter collapsing into a Schwarzschild BH does have the decay time scale governed by the Compton wavelength of the corresponding gauge boson. Thus, according to the results of \cite{Pawl}, the prediction of \cite{Coleman} is correct at least in monopole case.

In this paper we show that the time scale of the screening and the Proca hair loss is equal in order of magnitude to the gravitational radius of the BH and, therefore, the result \cite{Pawl} is incorrect. We analytically calculate the evolution of the field of the uniformly charged spherical shell surrounding the BH and contracting according to some undefined law. The results of this paper are valid for any classical Abelian massive vector field.

The study of the dynamics of the massive vector field in the vicinity of the BH, of course, is important for the phenomenology of quantum gravity, but may be more prolific cosmological consequences. 
An arbitrarily small but nonzero photon mass causes the effect of electrogenesis via BHs \cite{Dolgov2} with the rate of this process determined by the dynamics of the field created by the acreting charge. This process violates the electric symmetry of the Universe and probably makes a significant contribution to the generation of galactic magnetic fields. Analysis of the screening rate can allow us to establish the upper limit on the photon mass. The latter is important in the light of novel hypotheses, for example, explaining the origin of dark energy by introducing the mass of the electromagnetic field \cite{Kouwn}.

The massive photon can arise efficiently through the St\"uckelberg \cite{Ruegg} mechanism or spontaneous symmetry breaking via the Higgs mechanism. This implies the existence of a new elementary scalar field. However, the existence of the additional scalar field to produce the photon mass is not necessary. The photon mass, for example, can be generated topologically \cite{Dvali} by a four-dimensional analogue of the mechanism first described by Schwinger for two dimensions \cite{Schwinger}.

The article is organized as follows. In section \ref{sec2} we give a brief derivation of the static solution of the Proca equation for the fixed charged shell and discuss some related issues. 
The resolution of the dynamical problem is based on the static solution and described in detail in section \ref{sec3}. At the end of this section we give a formula to calculate the field evolution and define the bounds of its applicability. 
In section \ref{sec4} we make an attempt to discuss the possible limitations introduced by model imperfection.
In section \ref{sec5} we finally prove the discontinuity of the transition from the massive to the massless physics and give specific counterarguments to \cite{Pawl}. We conclude the article by summary of the results noting open areas for further research.

\section{Static solution of the Proca equation}\label{sec2}
Like any massive vector field generated by current density $j^\mu$, massive electromagnetic field with the photon mass $m$ satisfies the Proca equation
\begin{eqnarray}\label{Proca}
\nabla_\mu F^{\mu\nu}+m^2A^{\nu}=4\pi j^\nu,
\end{eqnarray}
where $F_{\mu\nu}=\partial_\mu A_\nu - \partial_\nu A_\mu$ and $\nabla$ denotes the covariant derivative in the background gravitational field. Strictly speaking, the equation (\ref{Proca}) is valid if the Proca field and its source does not change space-time geometry. We will indeed assume throughout the paper that the stress-energy tensor of the Proca field and the mass of charged matter are negligible in this sense. Therefore, the gravitational field of the BH will be described by the unperturbed Schwarzschild metric
\begin{eqnarray}
ds^2=h(r)\,dt^2-dr^2/h(r)-r^2d\theta^2-r^2\sin^2(\theta)d\phi^2,\\
h(r)=1-r_g/r,
\end{eqnarray}
where $r_g$ is the gravitational radius of the BH. The latter is assumed to be surrounded by a concentric uniformly charged spherical shell with a charge $Q$ and a fixed radius $r_s$. The corresponding current density has only a temporal component
\begin{eqnarray}
j^t=\frac{Q\delta(r-r_s)}{4\pi r_s^2}.
\end{eqnarray}
It is natural to expect that the field inherits the symmetry of the metric and the source. Therefore $A_\phi$ and $A_\theta$ are equal to zero and the equation (\ref{Proca}) reduces to a single equation for the electrostatic potential
\begin{eqnarray}\label{stat}
\frac{1}{r^2}\partial_r(r^2\partial_r A_t)-\frac{m^2}{h(r)}A_t=4\pi j^t,
\end{eqnarray}
which can be solved by conversion to a form of the Whittaker equation \cite{Bateman}.
The corresponding general homogeneous solution is a linear combination of functions
\begin{eqnarray}
\label{st.sol1} A_1(r)= h(r)\,e^{-m(r-r_g)}\Phi(mr_g/2+1,2;2m(r-r_g)),\\
\label{st.sol2} A_2(r)= h(r)\,e^{-m(r-r_g)}\Psi(mr_g/2+1,2;2m(r-r_g)),
\end{eqnarray}
where $\Phi$ and $\Psi$ denote confluent hypergeometric functions \cite{Bateman} of the first and second kind, respectively.

From the no-hair theorem we know that any non-trivial homogeneous solution of the equation (\ref{stat}) cannot simultaneously satisfy physically justified boundary conditions at the surface of BH and spatial infinity.
The physically reasonable boundary conditions require the finiteness of observed invariants of the field
\begin{eqnarray}
I_1 \equiv F_{\alpha\beta}F^{\alpha\beta}\quad\mbox{and}\quad I_2 \equiv A_\alpha A^\alpha,
\end{eqnarray}
included in the Proca energy-momentum tensor that has the form
\begin{eqnarray}\label{tensor}
T^{\mu\nu}=\frac{1}{16\pi}[4F_{\beta\alpha}g^{\beta\mu}F^{\nu\alpha}+I_1 g^{\mu\nu}]-\frac{m^2}{8\pi}[2A^\mu A^\nu - I_2 g^{\mu\nu}].
\end{eqnarray}
Indeed, contributions of the solution (\ref{st.sol1}) into the field strength $E \equiv \partial_r A_t = - \sqrt{I_1}$ and the invariant $I_2$ are finite at the horizon, but grow without bound at $r\to\infty$.
On the other hand, as we approach the gravitational radius, the contribution of the solution (\ref{st.sol2}) to the field strength diverges logarithmically and the invariant $I_2$ grows infinitely according to the law $1/h(r)$. Considered at spatial infinity this solution vanishes quasi exponentially. Consequently, only trivial static solution is possible if the radius of the charged shell approaches gravitational one.

Solutions inside and outside the shell are represented by the functions $A_1$ and $A_2$ respectively as follows from the behavior of these functions described above.
From junction conditions of these solutions, it follows that the radial component of the field strength has the form
\begin{eqnarray}\label{statsol}\fl
E_0(r)=-2Qm\,\Gamma(1+mr_g/2)\left[\theta(r-r_s)A_1(r_s)\partial_r A_2(r)+\theta(r_s-r)A_2(r_s)\partial_r A_1(r)\right],
\end{eqnarray}
where $ \Gamma $ denotes the Euler gamma function \cite{Bateman} and $ \theta $ is the Heaviside step function.
We omitted here the details of pretty standard calculations.

Now let the mass of the gauge boson be small so that we can neglect the product $mr_s$ and, of course, $mr_g$. Then the field strength being measured by a distant observer outside the shell will be determined by a simple expression
\begin{eqnarray}
E_0(r)=\frac{Q}{r^2}\frac{r_s-r_g}{r_s}e^{-mr}(1+mr).
\end{eqnarray}
Thus, if the photon has an arbitrarily small but nonzero mass, the effective charge and the field disappear as the shell adiabatically approaches the surface of BH.
However, it is well known that in the case when the photon mass is strictly zero, the screening of the charge does not occur and we would simply have $E_0(r)=Q/r^2$ everywhere outside the shell regardless of its radius.

The nature of this discontinuity was in a varying degree studied in the papers \cite{Teitelboim,Leaute,Dolgov,Vilenkin}. The massive vector field acts as a source of space-time curvature, that is expressed by the second term in the stress-energy tensor (\ref{tensor}). Since the Proca equation is not gauge-invariant, the vector potential $A^\mu$ becomes physically observable.
Therefore, the boundary conditions at the horizon for the massive electromagnetic field do not coincide with the conditions in Maxwell's theory. For any non-zero photon mass, we must additionally require the invariant $I_2$ to be finite. It is worth noting that this discontinuity in the behavior of the effective charge takes place only for BH, i.e. if the source of spherically symmetric gravitational field has the event horizon \cite{Leaute}.

\begin{figure}
  \centering
  \begin{minipage}[t]{0.493\linewidth}
    \includegraphics[width=1\textwidth]{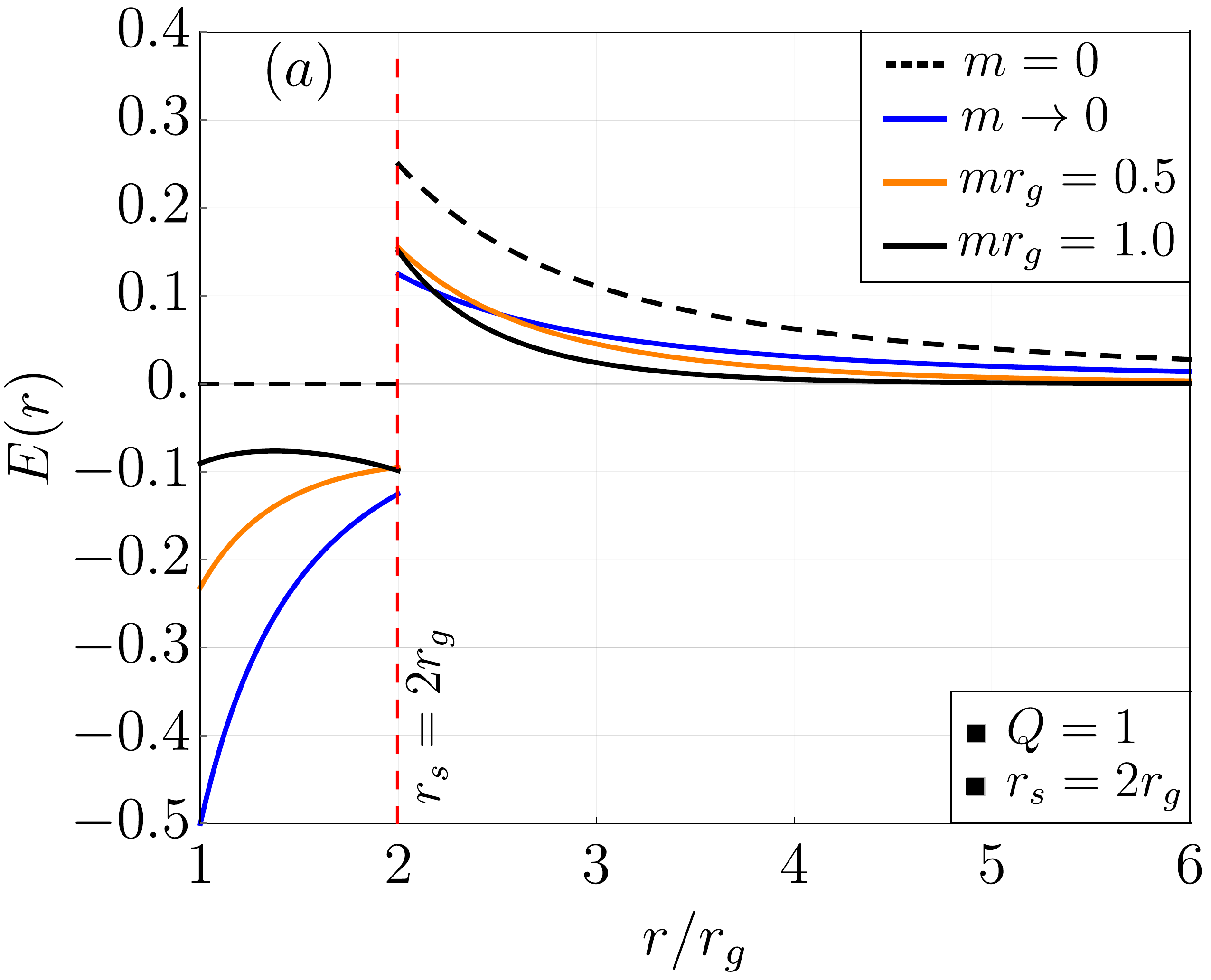}
  \end{minipage}
  \hfill
  \begin{minipage}[t]{0.493\linewidth}
    \includegraphics[width=1\textwidth]{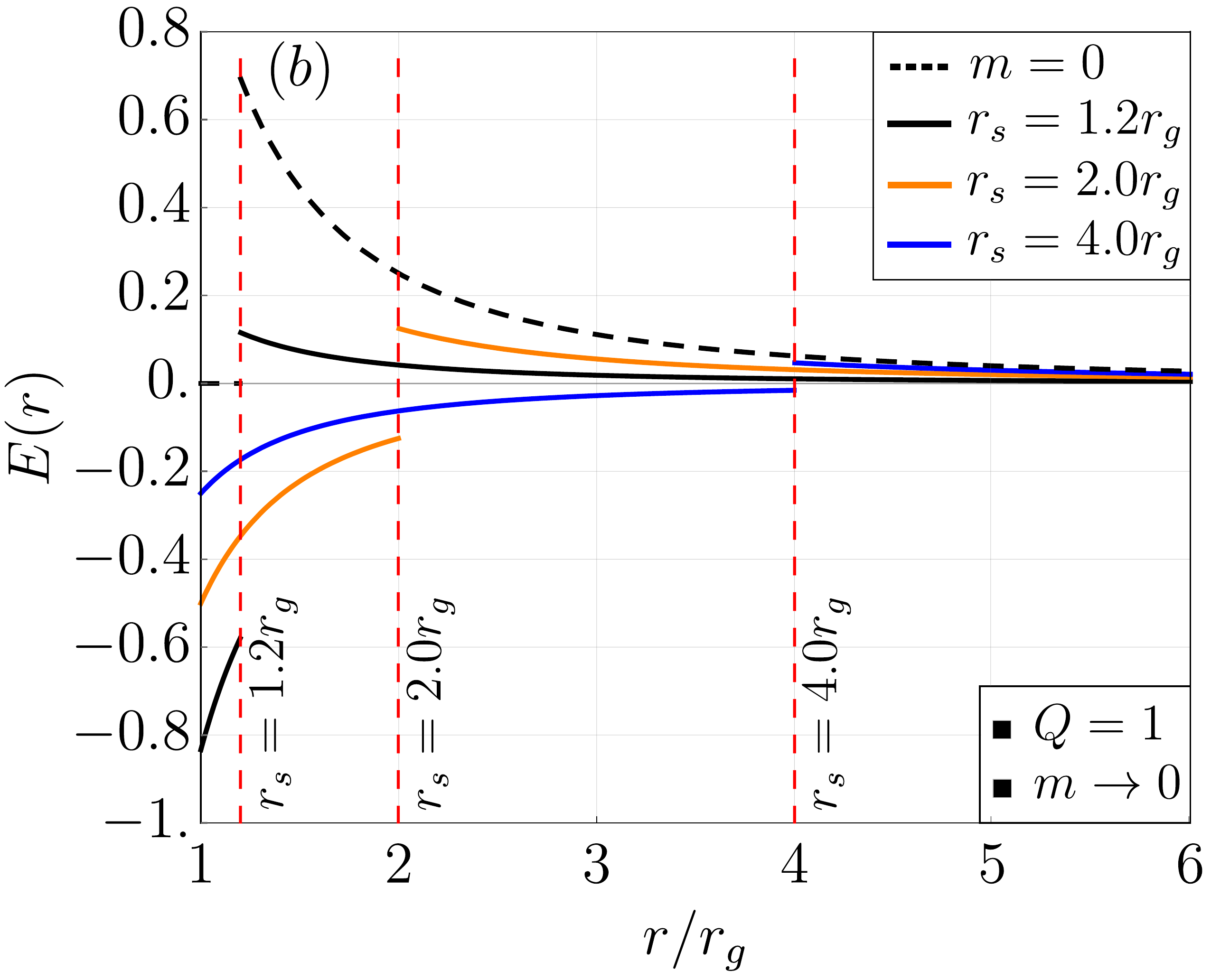}
  \end{minipage}
  \parbox[]{1\textwidth}{\caption{The strength $E=\partial_r A_t$ (in Schwarzschild coordinates) depending on the mass of the photon (a) and the position of the charged shell for the photon with a negligible mass (b). The additional parameters are also shown in the figure. Black dashed curves correspond to the Maxwell's case.}\label{fig1}}
\end{figure}

Shown in Figure \ref{fig1} is the behavior of the electrostatic field strength of a shell with a unit charge for various values of the photon mass and the shell radius in comparison with the massless case.

It is interesting also to note that the Gauss law, with some reservation, regains power in the limit of a negligibly small photon mass. If a charge $Q$ is contained in a volume $V$ (multiply connected, in general!) bounded by a surface $\partial V$ at some instant of the global time, the equation (\ref{Proca}) with $m \to 0$ gives
\begin{eqnarray}
\int_V \mathbf{\nabla} (\sqrt{-g}\,\mathbf{E})\,d^3 r = \oint_{\partial V} \mathbf{E} \,d^2 \mathbf{S} = \int_V \sqrt{-g}\,j^0\,d^3 r = 4 \pi Q,
\end{eqnarray}
where $\mathbf{\nabla}$ is an ordinary gradient and $\mathbf{E}$ is a strength vector in three-dimensional space. Generally speaking, this equality holds if the volume $V$ does not contain spatial projections of closed null hypersurfaces and, consequently, BHs. Since the latter capture a part of the electric field flux, as can be seen from the Figure \ref{fig1} (b), the Gauss law is generally violated.

\section{Analytical calculation of the field evolution}\label{sec3}
As we suppose now, the charged shell surrounding the BH begins to radially contract at some moment of the Schwarzschild time. We calculate the evolution of the Proca field associated with the shell, without specifying the driving force of the contraction and assuming that the stress-energy tensor of the source of this force is negligibly small. The dimensionless product $mr_g$ is also assumed to be small throughout this section.

Due to spherical symmetry, the field strength is connected to the vector potential by the relation $ E = \partial_t A_r- \partial_r A_t $. The Proca equation (\ref{Proca}) for this value has the form
\begin{eqnarray}\label{ref/1}
\frac{1}{h(r)}\partial_t^2 E - \partial_r\Big[\frac{h(r)}{r^2}\partial_r\left(r^2 E\right)\Big]+m^2 E=4\pi(\partial_t j_r-\partial_r j_t).
\end{eqnarray}
Keeping the notation introduced in the previous section the temporal and spatial components of the current density are determined as
\begin{eqnarray}
j_t=\frac{Qh(r)\delta(r-r_s)}{4\pi r_s^2} \quad \mbox{and} \quad j_r=-\frac{Q\partial_t r_s \delta(r-r_s)}{4\pi h(r) r_s^2}.
\end{eqnarray}

An analytical approach to calculating field behavior in this paper is based on the well-known method of retarded Green function. We apply this approach for the equation (\ref{ref/1}) previously translated into a frequency region by the Fourier transform, defined as follows:
\begin{eqnarray}
\mathcal{E}(\omega,r)=\int dt \, e^{-i \omega t}E(t,r),\\
E(t,r)=\frac{1}{2\pi}\int d\omega \, e^{i \omega t}\mathcal{E}(\omega,r).
\end{eqnarray}
The Green function $G=G(r,r_0|\omega)$ for the Fourier transform of the equation (\ref{ref/1}) by definition satisfies the equation
\begin{eqnarray}\label{GreenEq}
-\partial_r\Big[\frac{h(r)}{r^2}\partial_r(r^2 G)\Big]-\frac{\omega^2}{h(r)}G+m^2 G=\delta(r-r_0)
\end{eqnarray}
and the correct boundary conditions at the horizon which include the finiteness of the response function $G$. The homogeneous equation that corresponds to (\ref{GreenEq}) has the form
\begin{eqnarray}\label{homogeneous}
\partial_r^2\mathcal{E}+p(r)\partial_r\mathcal{E}+q(\omega,r)\mathcal{E}=0,
\end{eqnarray}
where we denote
\begin{eqnarray}
p(r)\equiv\frac{1}{r}+\frac{1}{r-r_g},\\
q(\omega,r)\equiv\frac{1}{r-r_g}\left[\frac{\omega^2 r^2}{r-r_g}-\frac{2}{r}\left(1-\frac{2r_g}{r}\right)-m^2 r\right],
\end{eqnarray}
and $\mathcal{E}$ is an unknown function to be determined. 

It is also useful to represent a homogeneous equation corresponding (\ref{ref/1}) by the standard replacement $E=\Psi/r$ and introducing the Regge-Wheeler coordinate
\begin{eqnarray}
r_\star=r+r_g \ln(r/r_g-1)+c,\nonumber
\end{eqnarray}
where $c$ is some constant. In this notation, the equation of the field evolution takes the form of the Schr\"odinger equation with an effective potential $V(r_\star)$:
\begin{eqnarray}
\label{Shredinger}\partial^2_t \Psi-\partial^2_{r_\star}\Psi+V(r_\star)\Psi=0,\\ \label{potential}
V(r_\star)=\left(1-\frac{r_g}{r}\right)\left[\frac{2}{r^2}+m^2-\frac{3r_g}{r^3}\right].
\end{eqnarray}
The frequency dependence of the transmission coefficient of the barrier formed by the effective potential $V$ determines the evolution of the field. The maximum of the barrier is located near $r=1.75\,r_g$ (for sufficiently small $m$).

We will mainly use the form (\ref{homogeneous}) throughout the rest part of the paper. A general homogeneous solution of this equation for $\omega = 0$ is obtained by taking the derivative of a linear combination of the functions (\ref{st.sol1}) and (\ref{st.sol2}) with respect to $r$. Solutions for the static problem in the regions $r<r_0$ and $r>r_0$ are respectively expressed as
\begin{eqnarray}\label{ref/3c1}
E_1(r)=\partial_r\left[h(r)e^{-m(r-r_g)}\,\Phi\left(mr_g/2+1,2;2m(r-r_g)\right)
\right],\\
\label{ref/3c2}
E_2(r)=\partial_r\left[h(r)e^{-m(r-r_g)}\,\Psi\left(mr_g/2+1,2;2m(r-r_g)\right)\right].
\end{eqnarray}
In order to find the solution in the case of an arbitrary law of the shell contraction, we will require also the smallness of the product $\omega r_g$ which is not a significant constraint of generality. 

We will start to solve the equation (\ref{homogeneous}) by computing the solution in the region with $r < r_0$.
Recall first the boundary conditions for the solution at the horizon. Being taken at the point $r = r_g$ the indicial equation $\rho(\rho-1) + \rho + r_g \omega = 0$ has the roots $ \rho = \pm ir_g \omega $. Since the difference of these roots does not equal to an integer real number for any nonzero product $\omega r_g$, both linearly independent solutions in a sufficiently small neighborhood of the gravitational radius can be found in the form of a series \cite{Whittaker}. Therefore, unlike the static problem, we cannot choose here the correct combination of the linearly independent solutions, relaying only on their finiteness in a neighborhood of the gravitational radius.

In order to determine a desired combination of the solutions, it will be useful to study its behavior in the limit $r_g\to 0$. It is natural to require that the solution in this limit becomes the same as for the case without the BH. Strictly speaking, this follows from the fact that in the case of flat space-time, due to spherical symmetry and vector nature of the field, the latter vanishes in the center of the charged sphere for any law of its contraction. Thus, the emergence of a sufficiently small BH in the center of the shell will not have any significant impact on the behavior of the field everywhere outside the BH, since the interaction between the field of the shell and the gravity of the BH in its exterior is negligible.
Introducing $\gamma=\sqrt{m^2-\omega^2}$ we can write the general homogeneous solution of the equation (\ref{homogeneous}) in flat space-time in the form
\begin{eqnarray}\label{ref/10}
\mathcal{E}_0(\omega,r)=\frac{m^2}{2\gamma^3 r^2}\left[\alpha\,e^{-\gamma r}(1+\gamma r)+\beta\,e^{\gamma r}(\gamma r-1)\right],
\end{eqnarray}
where $\alpha$ and $\beta$ are some constants. The only combination being regular in the center of the spherical coordinate system is obtained when these constants are equal to each other that is physically explained by vector character of the field and energy balance between converging and diverging waves. We will consider that $\alpha=\beta=1$ in the limit of vanishing curvature.

How will the solution change when the BH appears? It is natural to expect that the smallness of the parameters $mr_g$ and $\omega r_g$ implies a high quality approximation of the homogeneous solution by a superposition (\ref{ref/10}) in the limit of large values of the radial coordinate. But how will the balance between the coefficients $\alpha$ and $\beta$ change?
The answer can in fact be easily obtain directly from the properties of the expression (\ref{ref/10}).
Because $\partial_t g_{\mu\nu}=0$ in the Schwarzschild geometry, the energy is conserved that requires $|\alpha|=|\beta|$. Hence, any violation of this equality means that the field changes the metric. In the specific case, this change is explained as the wave absorption by the BH. The influence of this process on the field dynamics can be neglected, since the characteristic wavelengths, as we require, significantly exceed the gravitational radius. Thus, we can require that the amplitudes of the ingoing and outgoing waves coincide in absolute value.

Consider the dependence of the effective potential (\ref{potential}) on the radial coordinate. As can be seen, the influence of the mass term on behavior of the desired homogeneous solution can be neglected near the BH, while the terms with $r_g$ can be neglected for sufficiently large values of the radial coordinate where solution behavior significantly depends on $m$. A coordinate of the border between these regions can be defined by the equality in absolute value of the last two terms in square brackets, which gives
\begin{eqnarray}
r_b\equiv\sqrt[3]{\frac{3r_g}{m^2}}.
\end{eqnarray}
The regions with $r<r_b$ and $r>r_b$ will be referred to below as regions of dominant curvature and mass, respectively. As expected, this distinction is also justified by the behavior of the static solution (\ref{ref/3c1}), since its expansion has the form
\begin{eqnarray}\label{ref/decomposition}
E_1(r)=\frac{r_g}{r^2}+\frac{1}{3}m^2r + o(m^2r).
\end{eqnarray}
with the same terms in the same proportion. We will require the desired solution $\mathcal{E}$ to go into the static $E_1$ in the limit $\omega\to 0$.

Now we can simply solve our problem with stated accuracy if the region with $r<r_b$ is sufficiently wide that corresponds to the inequality $r_b \gg r_g$. Recall that the equations (\ref{homogeneous}) and (\ref{Shredinger}) are equivalent and $r_\star$ is asymptotically equals to $r$ for large $r/r_g$. The function describing field behavior in the mass dominated region far enough from its border is the flat space-time solution that we determine as the function (\ref{ref/10}) with $\alpha=\beta=1$:
\begin{eqnarray}\label{asym2}
\mathcal{E}_m(\omega,r)\equiv\frac{m^2}{2\gamma^3 r^2}\left[e^{-\gamma r}(1+\gamma r)+e^{\gamma r}(\gamma r-1)\right]=\frac{m^2}{\gamma^2}\partial_r\frac{\sinh{\gamma r}}{\gamma r}.
\end{eqnarray}

To compute the corresponding approximation to the solution in the curvature dominated region, we first assume that $r_g\ll r \ll r_b$.  The effective potential $V\sim 2/r^2$ in this spherical layer and thus we can neglect both the mass and the gravitational radius in the equation (\ref{homogeneous}) to find a function approximating the solution in the aforementioned layer using this equation. Of course, both the equations have the same general solution which is easy to derive. It remains only to choose the correct linear combination of the corresponding functions. Since the function (\ref{asym2}) behaves as $m^2r/3$ for small values of the product $m r$ and in the additional limit $\omega\to 0$, the desired form of the approximate solution for said range of $r$ in the low frequency limit goes to $r_g/r^2$, as seen from the expansion (\ref{ref/decomposition}). This property is inherent in only one combination of these functions that respects energy balance between converging and diverging waves and has the form  
\begin{eqnarray}\label{asym1}
\mathcal{E}_{r_g}(\omega,r)\equiv\frac{r_g}{2r^2}\left[e^{-i\omega r}(i\omega r+1)-e^{i\omega r}(i\omega r-1)\right]=-r_g\,\partial_r\frac{\cos{\omega r}}{r}.
\end{eqnarray}
In fact, this expression can be considered with good accuracy applicable up to the horizon. Indeed, the solutions of the equation (\ref{Shredinger}) in the frequency region and in the limit of large $\omega r_g$ are represented by a combination
\begin{eqnarray}\label{WKB}
\Psi(\omega,r)=A_\omega \exp{(i\omega r_\star)}+B_\omega \exp{(-i\omega r_\star)}+\mathcal{O}(\omega r_g)^{-1},
\end{eqnarray}
where $A_\omega$ and $B_\omega$ are coefficients. However this expression is valid for any frequency $\omega$ in close neighborhood of the horizon.
Since we require the product $\omega r_g$ to be small the product $\omega r$ with radii $r$ of the order of $r_g$ is also a small value. Hence, the constant component of the solution dominates near the BH except for a small neighborhood of the horizon, where it is necessary to take into account logarithmic divergent terms contained by (\ref{WKB}). The latter also means that the condition of smallness of the product $\omega r_g \ln(r/r_g-1)$ is a more precise applicability restriction. Thus, we have determined and joined the approximate solutions in the regions of dominant curvature and mass far enough from the border between these regions. 

The law of the shell contraction is required to be described by a smooth function decreasing monotonically. If $r_i$ and $r_f$ denote the initial and final radii of the shell, the Fourier image of the function in the right-hand side of the equation (\ref{ref/1}) equals to zero in the regions $r>r_i$ and $r<r_f$. We assume in this paper that $r_i \ll r_b$ and the observer measuring the Proca field is far from the source of gravity, so that its radial coordinate $r$ is large compared to $r_g$ and $r>r_i$. Given these assumptions, the Green function can be written in the form
\begin{eqnarray}\label{ref/green1}
G(r,r_0|\omega)=\mathcal{E}_{<}(\omega,r_0)\mathcal{E}_{>}(\omega,r)/\left[h(r_0)W(\mathcal{E}_{>}(\omega,r_0),\mathcal{E}_{<}(\omega,r_0))\right],
\end{eqnarray}
where the function $\mathcal{E}_<=\mathcal{E}_{r_g}$ is the homogeneous solution for $r<r_0$, $\mathcal{E}_>$ denotes the homogeneous solution in the range $r>r_0$ to be represented by its approximate form and $W(\mathcal{E}_>,\mathcal{E}_<)$ is the Wronskian of these functions. The function $\mathcal{E}_>$ describes a diverging wave in both the curvature and mass dominated regions as well as there is no parameter in the problem that depends on whether the coordinate $r$ belongs to a particular region. It follows that we can consider the form of the function $\mathcal{E}_>$ does not depend on the region. Therefore, we can fix the observer and study the limit of the Proca mass going to zero, when the region of dominant curvature expands unlimitedly.

The solutions $\mathcal{E}_<$ and $\mathcal{E}_>$ are well known from the theory of differential equations to be connected by the integral relation
\begin{eqnarray}\label{ref/connection}
\mathcal{E}_{>}(\omega,r)=\mathcal{E}_{<}(\omega,r)\int^r \frac{d\xi}{\xi(\xi-r_g)\mathcal{E}^{2}_{<}(\omega,\xi)},
\end{eqnarray}
with the integration constant determined for the solution $\mathcal{E}_>$ to vanish at spatial infinity. Given the asymptotic form of $\mathcal{E}_<$, the relation (\ref{ref/connection}) allows us to determine the asymptotic form of $\mathcal{E}_>$. To find this form in the frame of the assumptions made, it is enough to set $r_g = 0$ in this integral relation and use the expression (\ref{asym2}) as the asymptotic form of $\mathcal{E}_<$.
Substituting the expression (\ref{ref/connection}) into the formula (\ref{ref/green1}) we finally obtain a rather simple expression for the Green function:
\begin{eqnarray}\label{ref/green2}
G(r,r_0|\omega)=-r_0^2 \mathcal{E}_>(\omega,r)\mathcal{E}_<(\omega,r_0)
=-\partial_r\frac{e^{-\gamma r}}{m^2r}r_0^2 \mathcal{E}_<(\omega,r_0).
\end{eqnarray}

If the law of radial contraction of the shell is described by a smooth function of time, the Fourier image of the right-hand side of the equation (\ref{ref/1}) is represented as
\begin{eqnarray}
\label{ref/rhs1}
F(\omega,r)=Q\left[F_T(\omega,r)+F_R(\omega,r)\right],
\end{eqnarray}
where $F_T$ and $F_R$ denote contributions from the temporal and spatial components of the current density, namely
\begin{eqnarray}
\label{ref/rhs2}
F_T(\omega,r)=\partial_r\left[\partial_r t_s(r)e^{-i\omega t_s(r)}h(r)/r^2\right],\\
\label{ref/rhs3}
F_R(\omega,r)=\frac{i\omega e^{-i\omega t_s(r)}}{h(r)r^2}.
\end{eqnarray}
Here $t_s(r)$ is the function inverse to $r_s(t)$.

Therefore, the strength of the Proca field outside the shell at a large distance from the BH will be described by the expression
\begin{eqnarray}\label{ref/11}
E(\tau,r)=\int \frac{d\omega}{2\pi}e^{i\omega \tau}\int_{r_f}^{r_i} d\xi G(r,\xi|\omega)F(\omega,\xi)=\nonumber\\
-\partial_r\int\frac{d\omega}{2\pi}\frac{e^{i\omega\tau-\gamma r}}{m^2r}\int_{r_f}^{r_i}\xi^2 \mathcal{E}_<(\omega,\xi) F(\omega,\xi)d\xi,
\end{eqnarray}
where $\tau$ is the Schwarzchild time coordinate of the observer. Recall that the function $\mathcal{E}_<$ is entire for any finite $\xi>r_g$ in the frame of our approximation. Therefore, the Green function has only branch points corresponding $\gamma=0$.
To satisfy retarded boundary conditions, the contour of the integration goes around these singularities at Im\,$\omega<0$.

Of course, the expression (\ref{ref/rhs2}) is by no means regular function of $\omega$ in a neighborhood of zero frequency.
It is advisable to explicitly recover the corresponding singularity to take the inverse Fourier transform in the expression (\ref{ref/11}). In order to avoid using cumbersome expressions, we first introduce a notation
\begin{eqnarray}\label{func}
\mathcal{F}(\omega,r)\equiv\partial_r\left[\partial_r\left[r^2\mathcal{E}_<(\omega,r)\right]h(r)/r^2\right].
\end{eqnarray}
To regularize the integral over $\xi$, let also $\rho$ be the shell radius at some point in the time denoted by $t_0$. Integrating now twice in parts the relevant function of $\xi$ in the expression (\ref{ref/11}) from $r_f$ to $\rho$, we obtain    
\begin{eqnarray}\label{reg}
\mathcal{I}_1(r_f,\rho)\equiv\int_{r_f}^{\rho}d\xi \xi^2 \mathcal{E}_<(\omega,\xi)F_T(\omega,\xi)=e^{-ipt_0}\mathcal{E}_<(\omega,\rho)h(\rho)\partial_{\rho}t_s(\rho)\nonumber\\
+e^{-ipt_0}\partial_{\rho}[\rho^2\mathcal{E}_<(\omega,\rho)]\frac{h(\rho)}{i\rho^2p}-\frac{1}{p}\int_{r_f}^{\rho}d\xi e^{-ipt_s(\xi)}\mathcal{F}(\omega,\xi),
\end{eqnarray}
where $p=\omega-i\epsilon$ and $\epsilon$ is an infinitesimal positive parameter of regularization. The behavior of the field in the light cone of the future with the vertex at the point $(\rho,t_0)$ is not completely determined by this expression, in contrast to the case of a massless field. Since the propagation speed of a massive field is less than the speed of light, the behavior of the field in the space-time region mentioned above depends, strictly speaking, on the entire history of the change in the state of the shell at the time $t<t_0$. However, we can easy make the expression (\ref{reg}) to work taking $t_0$ to the distant past and requiring the product $\epsilon t_0$ to be small in this limit. Wherein the influence of the history of the shell until the time $t_0$ becomes negligibly small and this expression can be greatly simplified. Namely, in the subsequent integration over $\omega$ due to the infinitely fast oscillating behavior of the nonintegral terms, the contribution of the first term to the Fourier integral is zero, while the contribution of the second term can be calculated through a residue. Given the latter, we can make an equivalent replacement
\begin{eqnarray}
\mathcal{I}_1(r_f,r_i)\to 2\pi \delta(\omega) \partial_{r_i}[r_i^2\mathcal{E}_<(\omega,r_i)]\frac{h(r_i)}{r_i^2}
\label{ref/13}
-\frac{1}{i\omega}\int^{r_i}_{r_f}d\xi e^{-i\omega t_s(\xi)}\mathcal{F}(\omega,\xi),
\end{eqnarray}
where we imply that the pole in the second term belongs to the upper halfplane. As it should be, the contribution of this term into the integral over $\omega$ is equal to zero if the condition $\tau-t(\xi)<r$ is satisfied. Indeed, the integrand is an analytic function in the lower halfplane and decays exponentially on the lower semicircle.

To conveniently represent the contribution of the first term in the expression (\ref{ref/13}) to the field strength, we should use the fact that the static homogeneous Proca equation corresponding to the equation (\ref{stat}) gives a relation
\begin{eqnarray}\label{Simplification}
\partial_r(r^2E_1)=m^2r^2e^{-m(r-r_g)}\Phi(mr_g/2+1,2;2m(r-r_g)).
\end{eqnarray}
It is easy to see now that the result of the inverse Fourier transform of the first term in the approximation of a small $mr_i$ has the form
\begin{eqnarray}\label{ref/12}
E_0(r_i,r)=\frac{Q}{r^2}h(r_i)e^{-mr}(1+mr),
\end{eqnarray}
already known to be the solution of the static problem for the case of the shell with the radius $r_i$. Keeping the introduced notation we can write the expression (\ref{ref/11}) as 
\begin{eqnarray}
E(\tau,r)= E_0(r_i,r)+\frac{Q}{2\pi}\int^{r_i}_{r_f}d\xi\,\partial_r\int d\omega\,e^{i\omega(\tau-t_s(\xi))-\gamma r}
\frac{\mathcal{F}(\omega,\xi)}{i\omega m^2 r}\nonumber\\
\label{ref/14}
-\frac{Q}{2\pi}\int^{r_i}_{r_f}\frac{d\xi}{h(\xi)}\partial_r\partial^2_\tau\int d\omega\,e^{i\omega(\tau-t_s(\xi))-\gamma r}
\frac{\mathcal{E}_<(\omega,\xi)}{i\omega m^2 r}.
\end{eqnarray}
This expression is dramatically simplified, because using the equation (\ref{homogeneous}), the function $\mathcal{F}$ defined by the formula (\ref{func}) can easily be converted to the form
\begin{eqnarray}\label{relation}
\mathcal{F}(\omega,r)=m^2\mathcal{E}_<(\omega,r)-\frac{\omega^2}{h(r)}\mathcal{E}_<(\omega,r).
\end{eqnarray} 
Being applied in the expression (\ref{ref/14}) this relation generates a term that diverges in the limit $m\to 0$. But it is easy to see that the contribution from the last term in the right-hand side of the relation (\ref{relation}) completely reduces the last term in this expression. This reduction saves the only term that are finite in the limit of $m$ going to zero. The expression (\ref{ref/14}), up to negligible corrections, takes the form
\begin{eqnarray}
E(\tau,r)= E_0(r_i,r)+\frac{Q}{2\pi}\int_{r_f}^{r_i} d\xi\,\partial_r\int d\omega \,e^{i\omega \Delta\tau(\xi)-\gamma r}\frac{\mathcal{E}_<(\omega,\xi)}{i\omega r}\nonumber\\
=E_0(r_i,r)+\frac{Qr_g}{2\pi}\int_{r_f}^{r_i}\frac{d\xi}{\xi^2}\partial_r\int \frac{d\omega}{i\omega r} e^{i\omega\Delta\tau(\xi)-\gamma r}\cos{\omega\xi}\nonumber\\ \label{field}
-\frac{iQr_g}{2\pi}\int_{r_f}^{r_i}\frac{d\xi}{\xi}\partial_r\int \frac{d\omega}{r} e^{i\omega \Delta\tau(\xi)-\gamma r}\sin{\omega\xi},
\end{eqnarray}
where we define $\Delta\tau(\xi)\equiv\tau-t_s(\xi)$.
Therefore, to go from the frequency plane into the time region, we need to calculate the integral
\begin{eqnarray}\label{ref/integral}
\mathcal{I}_2(\alpha,\beta)\equiv\frac{1}{2\pi}\int \frac{d\xi}{i\xi} \exp(i \alpha \xi-\beta \sqrt{1-\xi^2}) = \nonumber\\ \int^\alpha_{-\infty}\frac{d\xi}{2\pi}\int d\eta \exp(i\xi\eta-\beta\sqrt{1-\eta^2}),
\end{eqnarray}
which depends on the parameters $\alpha$ and $\beta$. The choice of the lower limit in the right-hand side of this expression dictated by the fact that the initial integral is equal to zero if $\alpha=-\infty$ or, more precisely, if $\alpha<\beta$ since the singularities of the integrand are located above the real axis that corresponds to retarded boundary conditions.

The inner integral in the right-hand side of the expression (\ref{ref/integral}) is connected by a simple relation with the retarded Green function of the Klein-Gordon equation. This function is known to be written as
\begin{eqnarray}\label{ref/KFG}
D_{ret}(t,r)=\frac{1}{2\pi}\theta(t)\left[\delta(\lambda^2)-\theta(\lambda^2)\frac{m}{2\lambda}J_1(m\lambda)\right],
\end{eqnarray}
where $\lambda^2=t^2-r^2$ and $J_1$ is the Bessel function of the first kind. Indeed
\begin{eqnarray}\label{ref/18}
D_{ret}(t,r)=-\frac{1}{(2\pi)^4} \int \frac{d\omega\,d^3k\,\exp[i\omega t-i \mathbf{kx}]}{\omega^2-\mathbf{k}^2-m^2+i\epsilon\omega}=\nonumber\\
\frac{1}{8\pi^2 r}\int d\omega \exp(i\omega t-r\sqrt{m^2-\omega^2+i\epsilon\omega}),
\end{eqnarray}
that can be easily proved by integration over the components of the vector $\mathbf{k}$ in the four-dimensional integral. Taking into account the formulas (\ref{ref/KFG}) and (\ref{ref/18}) we obtain
\begin{eqnarray}\label{ref/integral2}
\mathcal{I}_2(\alpha,\beta)=\beta\int^\alpha_{-\infty}d\xi\,\theta(\xi)\left[2\delta(\lambda^2_\xi)-\theta(\lambda^2_\xi)J_1(\lambda_\xi)/\lambda_\xi\right],
\end{eqnarray}
where we denote $\lambda_\xi=\sqrt{\xi^2-\beta^2}$. Hence, we can write
\begin{eqnarray}
\label{ref/int1}
\partial_\beta[\mathcal{I}_2(\alpha,\beta)/\beta]=\partial_\beta[\mathcal{D}(\alpha,\beta)+e^{-\beta}/\beta]\,\theta(\lambda^2_\alpha)-2\theta(\alpha)\delta(\lambda^2_\alpha),\\
\label{ref/func}
\mathcal{D}(\alpha,\beta)\equiv\int^\infty_\alpha d\xi J_1(\lambda_\xi)/\lambda_\xi.
\end{eqnarray}

Combining the expressions (\ref{field}) and (\ref{ref/int1}) and getting rid of the generalized functions we eventually find the final somewhat cumbersome expressions describing the Proca field strength measured by a distant observer:
\begin{eqnarray}\label{res}
E(\tau,r)=\frac{1}{2}\left[E_0(r_{s+},r)+E_0(r_{s-},r)+R_+(\tau,r)+R_-(\tau,r)\right],\\
R_{\pm}(\tau,r)\equiv\frac{Qr_g \dot{r}_{s\pm}}{r r_{s\pm}^2} \pm \frac{Qr_g\dot{r}_{s\pm}}{r r_{s\pm}}\bigg[\frac{\dot{r}_{s\pm}}{r_{s\pm}}-\frac{\ddot{r}_{s\pm}}{\dot{r}_{s\pm}}-\frac{1}{r}\bigg]\nonumber\\
+Qmr_g\int^{r_i}_{r_{s\pm}}\frac{d\xi}{\xi^2}\partial_r\mathcal{D}(m\Delta\tau(\xi)\pm m\xi,mr)\nonumber\\ \mp Q m^2r_g\partial_r\int^{r_i}_{r_{s\pm}}\frac{d\xi}{\xi}\frac{J_1(m\sqrt{[\tau-t_s(\xi)\pm \xi]^2-r^2}\,)}{m\sqrt{[\tau-t_s(\xi)\pm \xi]^2-r^2}}.
\end{eqnarray}
Here $r_{s\pm}$ is the root of the equation $\tau-t_s(\xi)\pm\xi=r$ with respect to $\xi$ and the point denote derivative by time $\tau$.

Let us briefly recall and clarify the limits of applicability for the derived expressions. 
Despite the fact that the source of the field was assumed during the derivation to be in the region of dominant curvature, this constraint can be omitted in the case of a small $mr_i$ because this condition provides the trivial dynamics in the limit $r_g\to 0$. In addition to the requirement that $mr_i$ and $r_g/r$ are small, we also implied the smallness of the product $\omega r_g \ln(r_f/r_g-1)$ that imposes an insignificant constraint on the class of possible laws of the shell contraction.
Obviously, this neglect is justified if the characteristic time scale of the contraction is substantially larger than $r_g$ and $r_f$ is not too close to $r_g$. It is worth noting here that the formulas obtained can be significantly simplified in the case of a nonrelativistic contraction of the shell. Under this assumption we can neglect the highest powers of the product $\omega \xi$ in (\ref{field}) and find a simple expression 
\begin{eqnarray}\label{res2}
E(\tau,r)=E_0(r_s,r)+\frac{Q r_g \dot{r}_s}{r^2_s r}+Qmr_g\int^{r_i}_{r_s} \frac{d\xi}{\xi^2}\partial_r\mathcal{D}(m\Delta\tau(\xi),mr),
\end{eqnarray}
where $r_s$ this time is the root of the simple equation $\tau-t_s(\xi)=r$.
On the other hand, the solution describing the field outside the shell is also approximated up to terms of the order of $\omega r_g\ln(r/r_g-1)$, as can be seen from the asymptotics (\ref{WKB}). Therefore, our result seem to potentially lose accuracy in the region of very large, but noninteresting, values of the radial coordinate.

Now analyse the behavior of the various terms in the right-hand side of the formulas (\ref{res}) and (\ref{res2}). The function $E_0$ determines the usual Yukawa contribution suppressed by the factor $h$.
The nonintegral terms correspond to the impulse response of the field to an instantaneous perturbation, with this response propagating at the speed of light.
Stretching behind the latter is a low-frequency tail described by the integral terms. Because the amplitude of this tail is suppressed by a factor $m^2$, the field evolution does not depend on the mass of the field if this mass is very small. Of course, the time scale $\tau$ of setting the field configuration that corresponds to the final radius $r_f$ depends on the characteristic rapidity of the shell contraction and on the radius of the shell. But the time of the transient in the limit $m \to 0$ is bounded below by $\tau \sim r_g$ that is clear already from an elementary dimensional analysis.

\begin{figure}
  \centering
  \begin{minipage}[t]{0.496\linewidth}
    \includegraphics[width=1\textwidth]{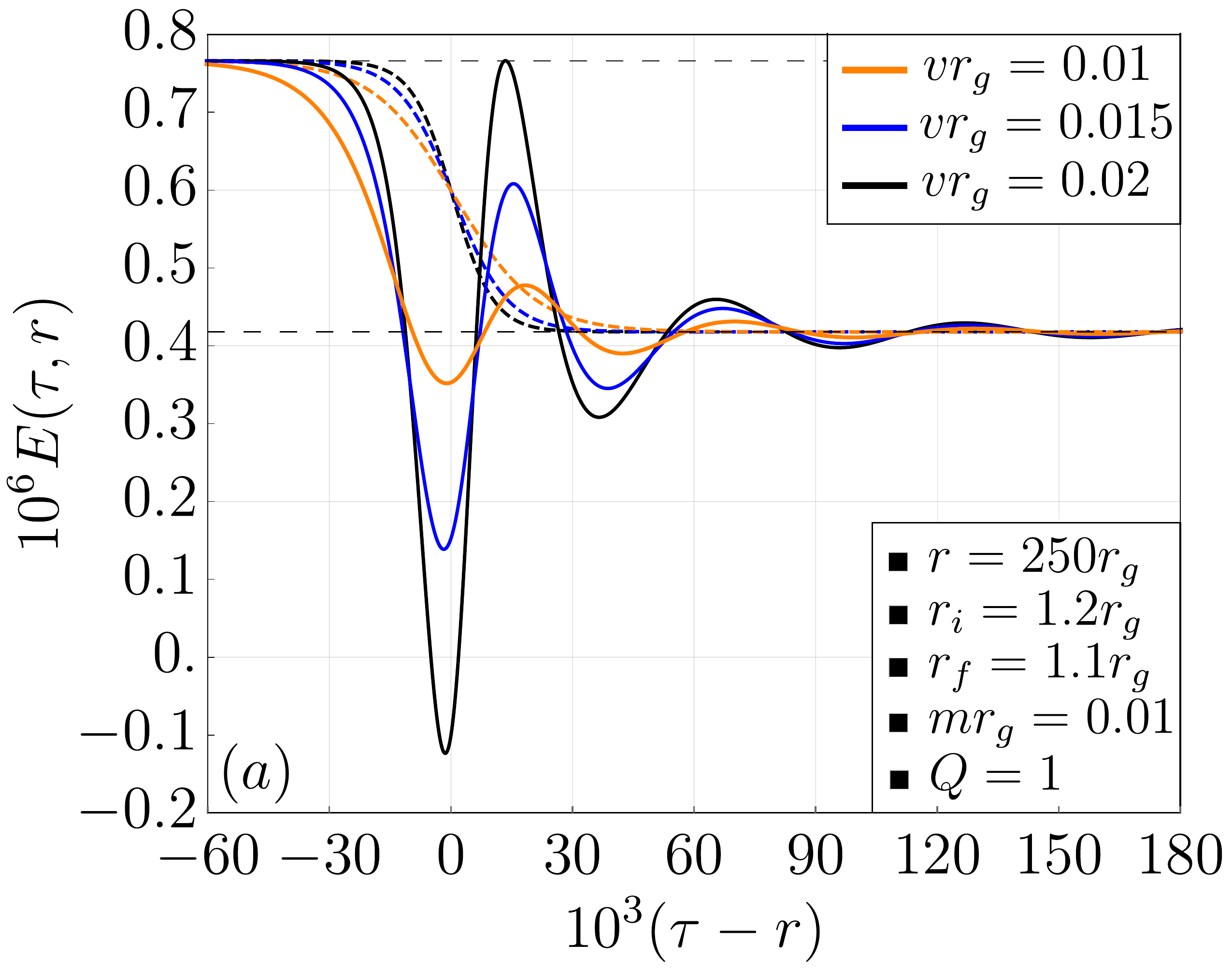}
  \end{minipage}
  \hfill
  \begin{minipage}[t]{0.496\linewidth}
    \includegraphics[width=1\textwidth]{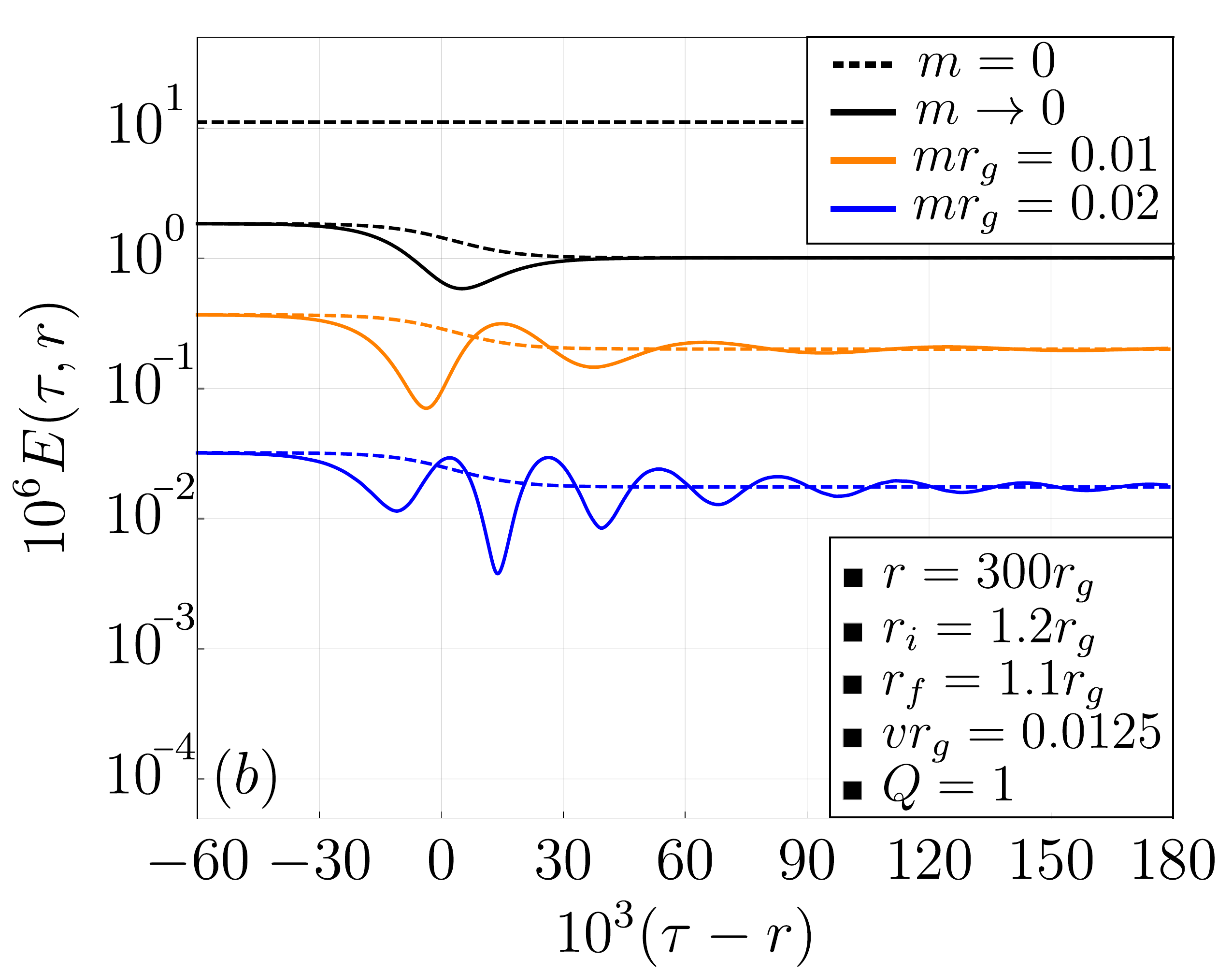}
  \end{minipage}
  \parbox[]{1\textwidth}{\caption{The evolution of the Proca field strength of the shell contracted according to the law (\ref{test}) depending on the rapidity (a) and the Proca mass (b). The additional parameters are also shown in the figure. Thin dashed curves correspond to the first term in the formula (\ref{res2}).}\label{fig2}}
\end{figure}

In order to demonstrate the typical behavior of the field, we accept, without further ado, the law of the shell contraction to have the form
\begin{eqnarray}\label{test}
r_s(t)=\frac{r_i+r_f e^{tv}}{1+e^{tv}},
\end{eqnarray}
where $v$ is the parameter setting the rapidity of the contraction. In accordance with the conditions of applicability, the product $vr_g$ must be small. 
Shown in Figure \ref{fig2} are temporal dependencies of the Proca field strength computed using the formula (\ref{res2}) for the law (\ref{test}) with different parameters $v$ and $m$.

Strictly speaking, the results obtained above are valid for the eternal BH and now we will discuss the influence of replacing this idealized source of gravity by a more realistic gravitating object.

\section{Notes on model validity}\label{sec4}

The key to uniquely determining the solution of the static problem and justifying the effect of electrical asymmetry in the case of the massive photon is divergence of the invariants $F^{\alpha\beta}F_{\alpha\beta}$ and $m^2A_\alpha A^\alpha$ in the stress-energy tensor (\ref{tensor}). However, as the mass of a vector field goes to zero, this divergence manifests itself in an increasingly narrow neighborhood of the horizon. The extremely low experimental upper bound for the photon mass, according to which $m<10^{-18}$ eV \cite{Tanabashi}, as well as the lack of full knowledge of the physical processes in the vicinity of the event horizon, may provoke a question whether a symmetry recovery mechanism exists for more realistic model of the BH, so that the symmetry will begin to recover for a certain value of the photon mass.

Suppose that in some $\delta$-neighborhood $U_\delta$:$(r_g,r_g+\delta)$ of the gravitational radius there are effects that regularize the invariants for both solutions $A_1(r)$ (\ref{st.sol1}) and $A_2(r)$ (\ref{st.sol2}), but leave these solutions practically unperturbed outside $U_\delta$. The corresponding source of gravity, generally speaking, is an exotic compact object (ECO) with "soft" hair \cite{Cardoso}, the radial coordinate of its surface belonging to $U_\delta$. One can represent up to a factor the electrostatic potential $A(r)$ in the region $r_g+\delta<r<r_s$ by a linear combination
\begin{eqnarray}
A(r)=A_1(r)+C(\delta)A_2(r)\nonumber\\
=\frac{r-r_g}{r}+C(\delta)\left[\frac{1}{2mr}+\frac{r-r_g}{r}\frac{\ln(2m(r-r_g))}{\Gamma(mr_g/2)}\right]+\mathcal{O}(mr),
\end{eqnarray}
with $C(0)=0$. The charge is not conserved if $C(\delta)=0$, but if $C(\delta)=2mr_g$ and $m$ goes to zero, the symmetry is restored, since the electrostatic potential inside the charged sphere of the radius $r_s$ is well known to be constant in the Maxwell case.

However, it is easy to conclude that the influence of processes that were not taken into account in the calculations for the case of the eternal BH does not cancel the electrical asymmetry in the limit $m\to 0$, if these effects are not connected with the destruction of the event horizon. Consequently, the existence of the horizon is a sufficient condition for the non-conservation of electric charge, not only in the idealized case.

Indeed, the solutions $A_1(r)$ and $A_2(r)$ under an idealized consideration have a clear physical sense which can be determined from their asymptotic behavior. The solution $A_2(r)$ describes the monopole field associated with a source fixed inside the sphere of the radius $r$, but outside the sphere of the radius $r_g$. In order to measure the magnitude of this field, we place a negligible test charge at a point with a radial coordinate $r$. Inevitable electromagnetic oscillations that accompany this process must also have negligible energy. But if the photon has a mass, the speed of wave propagation will depend on their frequency. Given the well-known properties of the event horizon, the interaction between the measured charge and the test one requires an unlimited increase in the energy of electromagnetic waves as the source of the measured field approaches the horizon. Therefore, the charge of a massive vector field that has crossed the horizon cannot manifest itself outside the BH. This situation does not occur for a strictly massless field.

Such an elementary interpretation explains the divergence of the invariants of the field energy-momentum tensor at the horizon for the solution $A_2(r)$. And this divergence is always present for at least one of the solutions of the static equation that describes the behavior of a massive vector field, regardless of the influence of any, so to speak, additional effects in the vicinity of the horizon.

One can consider that these effects include, for example, Hawking radiation. Evaporation of a BH by means of the latter or growth due to the effect of accretion of matter particles (ambient gas) and absorption of radiation, for macroscopic BH manifest themselves only in that the gravitational radius $r_g$ becomes time-dependent. Since these effects are not connected with the destruction of the event horizon (see, for example \cite{Unruh}), the qualitative picture of the evolution of the field in response to a change in the position of the charge remains undistorted. Quantitatively, the process of changing the radius of the horizon due to the mentioned processes, being slow compared to the characteristic relaxation time of the excitations of the Proca field and the background metric for macroscopic black holes, can be taken into account in the adiabatic approximation.

If the photon is the massive particle, the most general solution of the Einstein equation describing BHs is the Kerr metric. It is worth noting that BHs with nonzero angular momentum were found in \cite{Herdeiro} to be able to possess massive vector hair, but this hair has a harmonic temporal dependence and is not continuously connected with the Kerr-Newman solution. On the other hand, according to our results, it seems natural that the time scale of the screening of the field associated with a charge approaching the horizon is at least independent of the mass, so the transition from the massive case to the massless one is not continuous for all stationary BHs. The exact nature of the field behavior in the vicinity of the Kerr BH requires a separate study.

Finally, we should consider the influence of fact that BHs are formed as a result of the collapse. Evolution of the field in the exterior of a collapsing star requires a special consideration, which we will do in the next section.

\section{Gravitational collapse and decay of Proca hair}\label{sec5}

According to the Pawl results \cite{Pawl}, if a charged star collapses into a BH, an exterior massive vector field associated with a charge $q$ crossing the horizon fades over time $\tau_d$ of the order of $1/m$, with the Reissner-Nordstr\"om solution achieved continuously in the limit $m \to 0$ and parametrized by the same charge $q$. 

How then will the field evolve if the charge is outside the incipient BH? We assume again the charge forms a spherical shell of some fixed radius $r_s$ surrounding a \emph{neutral} star that at time $t=0$ begins to uniformly collapse into the BH with a radius $r_g$. If the spherical surface of last influence that narrows over time coincides with the shell at time $t_i>0$, any signal transmitted from the surface of the shell at the time $t>t_i$ is not able to reach the surface of the star until it completely hides under the horizon. This means that, from the point of view of an observer on the shell, the BH is already formed at the time $t_i$ and thereby we have exactly the same boundary conditions that were used in the derivation of our main result (\ref{res}). Therefore, the field configuration described by the expression (\ref{statsol}) is set during the time $\tau_s$ of the order of $r_g$ after $t_i$, because the field associated with the shell before $t_i$ obviously also disappears quickly being not connected with the charge crossing the horizon. Thus, taking into account the extreme smallness of $m$ in the hypothesis of massive photon, the effective charge of the shell experiences suppression much faster than the complete disappearance of the field associated with the charge absorbed by the BH in the process of its formation.

It may seem that such a difference in time scales can take place, because the problem of the author \cite{Pawl} is still different from the present one. But this is actually not the case. The event horizon is well known to form even before the stellar surface hides under the horizon \cite{Novikov}. The region of space-time hidden under the horizon expands with the speed of light until the moment when the surface of the star crosses the boundary of this region. A characteristic time scale of this process is consistent with $r_g$. Consequently, the exterior field created by the charge that is located in the volume of the star should also decrease quickly as the charge crosses the expanding horizon, since the collapse of the star is much faster process than the contraction of the charged shell considered in our problem. Recall that we require the characteristic time scale of the contraction to be large in comparison to the product $r_g \ln(r/r_g-1)$. Hence, we should consider that $\tau_d$ is of the order of $r_g$, which contradicts \cite{Pawl}.

Pawl used both analytical and numerical approaches to calculations in independent way. The analytical approach is based on a technique used by Price \cite{Price} to analyse the evolution of perturbations of a scalar massless field during the collapse of a star that contains matter acting as the source of this field.

To shed some light on the temporal dependence of the field created by a collapsing star and find the reason for the discrepancy with the result of the analytical approach, we will use the homogeneous equation (\ref{Shredinger}). 
We assume the collapsing star to be uniformly charged and this uniformity persists during the collapse. The effective potential (\ref{potential}) forms a barrier that completely reflects waves with the frequency going to zero. The frequency dependence of the transmission coefficient of this barrier determines the evolution of the field created by the star. Before the collapse, the field $\Psi$ outside the star of the radius $r_{sf}>r_g$ corresponds to the static solution of the equation (\ref{Shredinger}). If the total charge of the star is $Q$ and the mass of the photon can be neglected, then $\Psi=Q/r$ everywhere outside the star at the moment when the collapse begins.

An asymptotic law of temporal dependence of the field at the stellar surface while it approaches the horizon takes the form
\begin{eqnarray}\label{dilation}
\Psi(t,r_{sf})=a+b\exp\left[-\frac{t-r_\star(r_{sf})}{2r_g}\right],
\end{eqnarray}
where $a$ and $b$ are some constants. This law is a simple manifestation of the effect of time dilation and is the same for fields of any nature that have no pathological behavior in a neighborhood of the horizon. Its rationale for the example of a scalar massless field can be found in \cite{Price}. In the process of a collapse the expression (\ref{dilation}), as a rule, has a rather sharp time character leading to a pronounced presence of waves with frequencies 
\begin{eqnarray}
\omega\gg\sqrt{\max{V(r_\star)}}\sim 1/r_g,\nonumber
\end{eqnarray}
which pass through the potential barrier almost undistorted and thereby translate the exponential law (\ref{dilation}) of the field fading outside the barrier. Nevertheless, even before the time dilation begins to dominate, the field outside the star falling under the horizon will experience suppression due to the emission of waves of lower frequencies. The exact shape of the wavefront depends on the dynamics of the collapse.

Despite the fact this sketch of the picture of the field evolution in the vicinity of a nascent black hole was described by Price, the author of \cite{Pawl} limited himself to studying the asymptotic behavior of the field in the limit $t\to\infty$, which essentially boiled down to a simple analysis of the effective potential in the equation (\ref{Shredinger}) and led to a correct estimate of the time scale of the field change, but an incorrect interpretation of this result. The asymptotic behavior of a transient described by the formulas (\ref{res}) and (\ref{res2}) is indeed a superposition of oscillations with frequencies of the order of $m$, but its amplitude was already noted to be suppressed by the factor $m^2$ and therefore disappears in the limit of $m$ going to zero. This ringing effect at the final stage of the transient has nothing to do with the screening effect. But the behavior given by (\ref{res}) and (\ref{res2}) completely fits into the frame of the qualitative picture described above. Since Pawl studied only the temporal dependence, his analytical approach has no probative value and leads to an incorrect conclusion.

The numerical method \cite{Pawl} is criticized somewhat more simply. This time, the author mistakenly claims that the field strength in the immediate vicinity of the horizon tends to $Q/r^2$ in the limit $m \to 0$ at the time of the collapse completion and thus the formation of a BH. The continuous dependences of the equation (\ref{Shredinger}) and the initial conditions on the small parameter $mr_g$ is not enough to assert that the dynamics of the massive field during the collapse will be the same as in the massless case. It is necessary to analyse the behavior of the boundary conditions at the emerging horizon, but the author neglected this. As our computations have already shown, these conditions are discontinuous in the limit of $m$ going to zero.

Thus, the time scale $\tau_d$ of the decay of Proca hair in the process of a spherically-symmetric stellar collapse coincides in order of magnitude with $r_g$. The same time scale is necessary to consider the expression (\ref{res}) to be applicable. The connection between the massive case and the massless one is discontinuous, contrary to the statements of \cite{Coleman, Pawl}.

\section{Conclusion}

As was shown in this paper, the time scale of the screening of the Proca field associated with a source approaching the event horizon is determined by the mass of the BH. Being a conceptually identical process, the Proca hair loss during the collapse of a star has the same temporal behavior. Considered in the immediate vicinity of the event horizon, the physics of a classical massive vector field in the limit of the mass going to zero does not coincide with Maxwell’s physics.

The incorrect estimate of the time scale for the Proca hair to decay in the paper mentioned above led authors of \cite{Dolgov2} 
to a significantly low estimate of the electrical asymmetry of the Universe, which is generated by superheavy BHs in the centers of large galaxies. The difference in the mobilities of electrons and protons in the ionized plasma surrounding such BHs leads to an uncompensated radial current. Due to the radial redistribution of the charge, the positive charge of the protons is screened more strongly than the negative charge of the electrons, which leads to an excess of the negative charge in the Universe. A study of the generated asymmetry requires a careful approach and will be done in the next paper.

A sharp difference in the behavior of the low-frequency modes of electromagnetic radiation in the vicinity of a BH for massive photon and massless one provides a unique experimental opportunity to establish whether the photon has a mass or is it a massless particle \cite{Cameron}. A similar conclusion can be also made by comparing the theoretical and experimental \cite{Caprini} data on the electrical asymmetry of the Universe.

\ack
I wish to thank A. D. Dolgov for suggesting this problem. I grateful to A. A. Pomeransky and V. M. Khatsymovsky for the discussion and careful reading of the preprint. I also wish to thank A. S. Rudenko and A. V. Reznichenko for the helpful discussions. 
The work is supported by the RSF Grant No. 19-42-02004.


\end{document}